\documentclass[aps,twocolumn,pra,epsfig,showpacs]{revtex4}
\usepackage{graphicx}
\usepackage{epsfig}

\begin{document}

\title{Triatomic continuum resonances for large negative scattering
lengths}
\author{F. Bringas$^{(a)}$, M. T. Yamashita$^{(a)}$ and T. Frederico$^{(b)}$}
\affiliation{$^{(a)}$Laborat\'orio do Acelerador Linear, Instituto de F\'\i sica
da USP 05315-970, S\~ao Paulo, Brasil}
\affiliation{$^{(b)}$Dep. de F\'\i sica, Instituto Tecnol\'ogico de
Aeron\'autica, Centro T\'ecnico Aeroespacial, 12228-900 S\~ao
Jos\'e dos Campos, Brasil}
\date{\today}

\begin{abstract}
We study triatomic systems in the regime of large negative
scattering lengths which may be more favorable for the formation of
condensed trimers in trapped ultracold monoatomic gases as the
competition with the weakly bound dimers is absent. The
manipulation of the scattering length can turn an excited weakly
bound Efimov trimer into a continuum resonance. Its energy and
width are described by universal scaling functions written in
terms of the scattering length and the binding energy, $B_3$, of
the shallowest triatomic molecule. For $a^{-1}<-0.0297
\sqrt{m~B_3/\hbar^2}$ the excited Efimov state turns into a
continuum resonance.

\end{abstract}


\maketitle

It is by now well established that shallow dimers are formed in
trapped ultracold or condensed monoatomic gases, as it has been
reported for $^{23}$Na~\cite{mckenzie}, $^{87}$Rb~\cite{wynar},
and $^{85}$Rb~\cite{donley}. In the experiment in Ref.~\cite{donley}, 
an atom-molecule coherence in the Bose-Einstein condensate
was also observed. The measured oscillation frequency of the quantum
superposition of $^{85}$Rb dimers and atoms in the condensate was
in agreement with the shallow $^{85}$Rb$_2$ binding energy over a
wide range of values near a Feshbach resonance. However, the
formation of triatomic molecules has not yet been observed.

Recently, the energy of the shallowest bound triatomic molecule in
trapped ultracold and condensed monoatomic gases was predicted
using the value of the measured recombination rate into a shallow
dimer; the energies ranged from 7.75 mK down to 0.24 nK near a
Feshbach resonance~\cite{recomb}. Condensed triatomic molecules coexisting
with dimers and atoms near a Feshbach resonance would present an
oscillatory dependence of observables on the trimer binding
energy~\cite{braatenprl03}. For
negative scattering lengths and zero energy, the recombination
rate into deep diatomic molecular states shows a resonant peak at
values of $a$ for which the trimer Efimov state~\cite{ef70} hits
the three-body continuum threshold~\cite{braatenprl01}. If the
magnitude of the large negative scattering length is decreased
after an Efimov state hits the continuum, it turns into a
three-body resonance, as we will show. Therefore, at non zero
energies, or temperatures, the resonant peak of the recombination
rate would in principle appears when the resonance energy matches
the energy of the continuum triatomic system.

Furthermore, the manipulation of the scattering length in
monoatomic condensates in the regime of large and negative values
offers an interesting possibility. No shallow dimers can exist in
this case. Then, the formation of triatomic molecules may be more
favorable as the main competitors are absent. If a coherent
quantum superposition of atoms and trimers or resonances appears,
it will present an oscillatory frequency corresponding only to the
shallow three-body bound or resonant state. As we will show, this
frequency scales with the scattering length and with a triatomic
physical scale in the form of a universal function. Here, the
binding energy of the shallowest trimer state is chosen as the
three-body physical scale.

In 1970, Efimov predicted that infinitely many weakly bound
three-boson states appear when the $s-$wave two-boson scattering
lengths, $a$, goes to the limit of $a\rightarrow
\pm\infty$~\cite{ef70}. For large scattering lengths, an
attractive long-ranged effective interaction binds the
three-particle system in a range of about $|a|$ (See
also~\cite{jensen}). A new bound state appears for every increase
of $|a|$ to $\sim 23|a|$. For positive $a$, at the threshold of
the new state, the trimer binding energy is
$6.9B_2$~\cite{am99,yama02,braaten}($B_2$ is the dimer binding
energy). For $a<0$, a new state becomes bound when the trimer has
an energy of $\sim 1100B_2$~\cite{am99}, where now $B_2$ is the
energy of the virtual dimer state. Due to the large size of the
system, the threshold conditions for the existence of excited
Efimov trimers are universal~\cite{jensen,de00}, i.e., independent
of the detailed potential shape, and exhibit a scaling form for
large values of $a/r_0$~\cite{ef70,am99,braaten} (the interaction
range is $r_0$). The three-body system heals through regions that
are outside the potential action, where the wave function is
essentially a solution of the free Schr\"odinger equation, and
therefore the properties of the system are defined by few physical
scales.

The dimer and trimer binding energies are the only physical scales
that survive in the limit of $(a/r_0)\rightarrow \pm \infty$
(scaling limit), which essentially relates the Thomas collapse of
the trimer state for $r_0\rightarrow 0$~\cite{th35} to the Efimov
effect ($|a|\rightarrow \infty$)~\cite{adh88}. In the scaling
limit, the three-boson observables are functions of the
shallowest trimer binding energy (reference three-body energy) 
and $B_2$. These functions approach universal curves~\cite{ef70,am99}.
The collapse of the three-boson system in the limit of a
zero-range force makes the trimer energy  the
three-body scale of the system beyond the
two-body energy~\cite{ad95}.

For large scattering lengths, an excited Efimov state turns into a
virtual state when $a>0$ is decreased~\cite{tomio80}. The
threshold moves faster than the energy of the excited state.
Therefore, with the increase of $a>0$, states pump out from the
second sheet of energy to become bound states~\cite{yama02}. If
$a<0$ is decreased in magnitude, the trimer bound states dive into
the continuum~\cite{am99}. It is our aim here to evaluate the
scaling properties of the energy and width (for the decay into the
three-body continuum) of the resonance born from an Efimov state,
when a large $a<0$ is varied.

The $^4$He excited trimer state calculated also with realistic
models offers a good example of an Efimov state and its universal
scaling properties with the shallow dimer and trimer binding
energies~\cite{am99,braaten}. These molecules are special due to
the large spatial size which spreads out much beyond the range of
the potential~\cite{kiev01,roud00}. The $^4$He-$^4$He
root-mean-square distance in $^4$He$_3$ for the ground and excited
states~\cite{kiev01,moleculas} are of the order of 5 to 10\ \AA \
and of about 50 to 90\ \AA, respectively. The product of the
mean-square interatom distance with the separation energy of one
atom  from the trimer in units of $\hbar=m=1$ ($m$ is the atom
mass) is about the 1 in the ground and the excited
states~\cite{kiev01,moleculas}, while the ratio of the binding
energies $B^{(0)}_3/B^{(1)}_3\approx 500$~\cite{ef70} ($B^{(N)}_3$
is the binding energy of the $N^{th}$ trimer state).

In the present work we calculate the three-boson resonance energy
and width for the decay into continuum states for large and
negative scattering lengths. In this case it is justifiable to use
a Dirac-$\delta$ potential. We solve subtracted homogeneous
equations defined within a renormalization scheme applied to a
three-body system interacting with $s-$wave zero-range pairwise
potentials~\cite{ad95,yama02}. We present the results for the
energies and widths in the form of a universal scaling
function~\cite{am99}, which gives the trajectory of the three-body
energy in the complex plane as a function of the shallow two-body
virtual state energy. We show that a resonance becomes an excited
trimer Efimov state when the large $a<0$ is decreased. We go
beyond Ref.~\cite{yama02}, where it was found the dependence of the
virtual three-boson state energy with $a>0$ born from an Efimov
state which entered in the second energy sheet.

One can fix  the energy of one three-body bound-state (the
three-body physical scale), and the two-body scattering length and
get other observables. All the detailed information about the
short-range force, beyond the low-energy two-body observables, is
retained in only one three-body physical information in the limit
of zero-range interaction. The existence of a three-body scale
implies in the low energy universality found in three-body
systems, or correlations between three-body
observables~\cite{fre87a,ad95}. In the scaling
limit~\cite{am99,yama02}, one has
\begin{equation}
{\cal{O}}\left(E, B_{3},B_2\right)(B_{3})^{-\eta} = {\cal
F}\left(\sqrt{E/B_{3}},\pm\sqrt{B_2/B_{3}}\right), \label{o}
\end{equation}
where $\cal O$ is a general observable of the three-body system at
an energy $E$, with dimension of energy to the power $\eta$. This
equation means that any observable of the system can be
represented by a function that depends only on one three- and one
two-body scales. The three-body scale is brought by the reference
energy $B_3$ (binding energy) of the shallowest trimer state. The
$\pm$ sign denotes positive or negative scattering lengths. In the
case of the energies of a resonance, of an excited or
virtual trimer states, instead of Eq. (\ref{o}), the scaling
function is written as:
\begin{equation}
E_{3} = B_{3}~{\cal E}\left(\pm\sqrt{B_2/B_{3}}\right) \ .
\label{oe}
\end{equation}
(Throughout this paper, we use units such that $\hbar=m=1$, where
$m$ is the mass of the atom.)

After partial wave projection, the $s-$wave subtracted integral
equation for three identical bosons is given by~\cite{yama02}:
\begin{eqnarray}
 &&\chi(q)=4\pi\,\tau(\xi) \int_0^\infty dq^\prime {q^\prime}^2
\int_{-1}^1 dy \, \chi(q^\prime) \label{chi}
\\ \nonumber &&\times
\left(\frac{1}{E_3-q^2-{q^\prime}^2-q\,q^\prime \,y}-
\frac{1}{-\mu^2-q^2-{q^\prime}^2-q\,q^\prime \, y}\right) ,
\end{eqnarray}
where $\xi=E_3-\frac34 q^2$, $\mu$
is the subtraction point and $\tau$ is given by
\begin{eqnarray}
 \tau^{-1}(\xi)&=&-2\pi^2\,\sqrt{B_2} -4\pi\, \xi
 \int_0^\infty \frac{dp}{\xi-p^2}\ ,
\label{Tauq}
\end{eqnarray}
here $B_2$ is the dimer virtual state energy. (For positive values
of the real part of $E_3$, Eqs. (\ref{chi}) and (\ref{Tauq}) are
analytically extended to the second energy sheet, as we discuss
below.)

Throughout this work we perform calculations only considering
$a<0$ for a virtual dimer state. We use a contour deformation
method to calculate the resonance energy and width~\cite{complex}.
The homogeneous equation, (\ref{chi}), is analytically continued to the
second sheet of energy, by making
$q~(q^\prime)\rightarrow~q~e^{-\imath\theta}~(q^\prime
e^{-\imath\theta})$ with $0<\theta<\pi/4$. For large enough
$\theta$, the solution of Eq.(\ref{chi}) in the complex energy
plane is found for $\tan (2\theta)>-Im(E_3)/Re(E_3)$.

In the limit of a virtual dimer energy tending to zero an infinite
number of Efimov states appears from the solution of
Eq.~(\ref{chi}) for negative $E_3~(=-B_3)$. The $N^{th}$ Efimov
state has binding energy given by $B_3^{(N)}$ with $N=0$
indicating the ground state obtained from Eq.~(\ref{chi}). When
the $N^{th}$ Efimov state turns into a resonance its complex
energy is denoted by $E_3^{(N)}$.
\begin{figure}[htb]
\centerline{\epsfig{figure=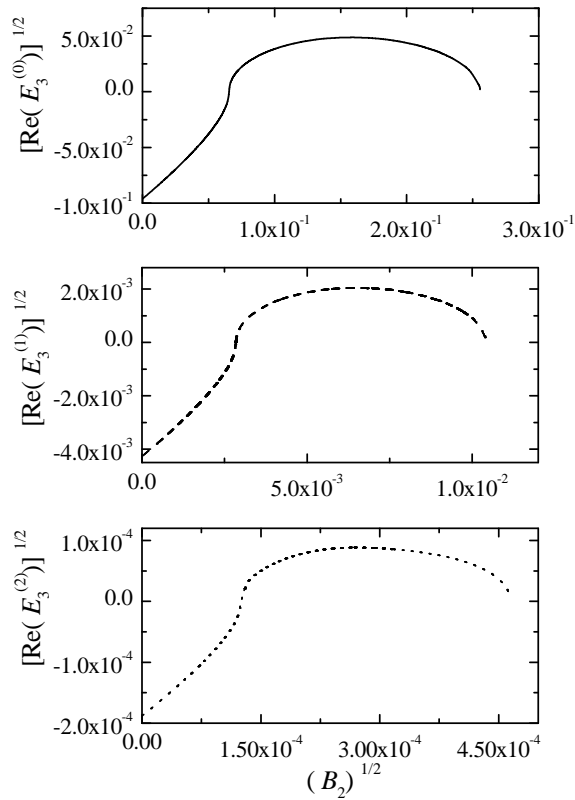,width=7.5cm}}
\caption[dummy0]{Real part of the three-body resonance energy as a
function of the two-body virtual state energy in units of $\mu=1$.
The negative values represents the results for square-root of the
energy of the $N^{th}$ bound trimer state, $[B_3^{(N)}]^{1/2}$,
which turns into a resonance when $|B_2|$ is increased. The
positive values of $[Re(E^{(N)}_3]^{1/2}$ corresponds to the
resonance. The results are given for $N=0$ (solid line), $N=1$
(dashed line) and $N=2$ (dotted line). } \label{fig1}
\end{figure}
In figure (\ref{fig1}), it is shown the real part of the complex
energies, in units of $\mu=1$, for the first three states obtained
by solving numerically Eq.~(\ref{chi}). In the figure, the values
of $[Re(E_3)]^{1/2}$ for the first three resonances are described
by the positive part of the plot. The curves represent $E_3^{(0)}$
(solid line), $E_3^{(1)}$ (dashed line) and $E_3^{(2)}$ (dotted
line), i.e., the complex energies of the first three states of
Eq.~(\ref{chi}). Note that the subtraction present in
Eq.~(\ref{chi}) regularizes it, and the Thomas collapse is
avoided. When $B_2$ is decreased the resonance turns into a bound
state and the negative part of the plot give the energies of the
ground ($B_3^{(0)}$), first ($B_3^{(1)}$) and second ($B_3^{(2)}$)
bound trimer states. One realizes that the form of the curves are
very similar which indicates that the function ${\cal E}$ of
Eq.~(\ref{oe}) does not depend on the state, which will be
confirmed later on, in the scaling plot of figure (\ref{fig3}).
The values of $\sqrt{B_2}$ in units of $\mu=1$ at which the
Efimov-state becomes unbound are 0.066, 0.0028 and 0.00013 for the
ground, first and second states, respectively. The ratios
$(0.066/0.0028)^2\approx (0.0028/0.00013)^2 \sim 500$ are
practically independent of the state, in agreement with the
scaling limit implied by Eq.~(\ref{oe}). It is curious that the
real part of the resonance energy tends to zero for $B_2$ large
enough, while the width is nonzero.

\begin{figure}[htb]
\centerline{\epsfig{figure=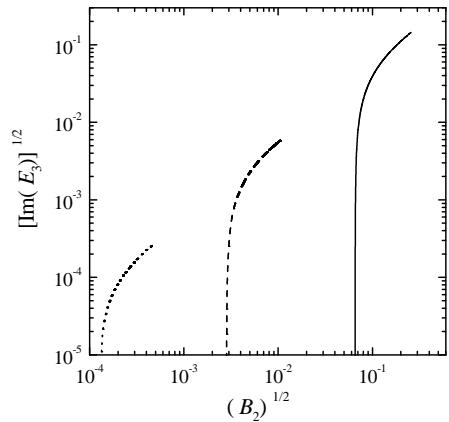,width=5.8cm}}
\caption[dummy0]{Imaginary part of the three-body resonance energy
as a function of the two-body virtual state energy in units of
$\mu=1$. Curves labelled as in fig.1~. } \label{fig2}
\end{figure}

\begin{figure}[htb]
\centerline{\epsfig{figure=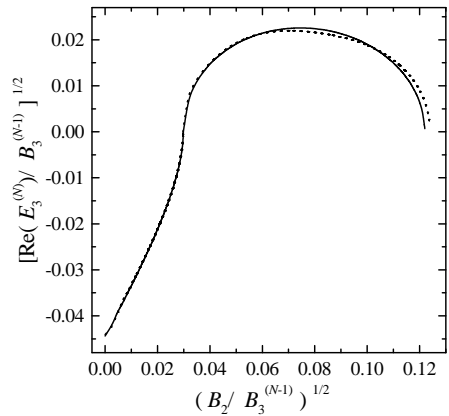,width=5.8cm}}
\caption[dummy0]{Ratio of the real part of $E_3^{(N)}$ and
$B_3^{(N-1)}$ as a function of $B_2/B_3^{(N-1)}$. The solid line
is the results for N=1 and the dashed line for N=2.} \label{fig3}
\end{figure}

\begin{figure}[htb]
\centerline{\epsfig{figure=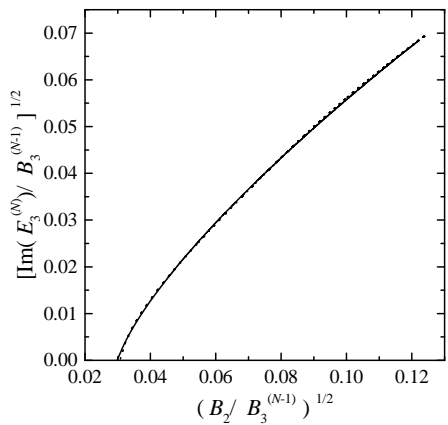,width=5.8cm}}
\caption[dummy0]{Ratio of the imaginary part of $E_3^{(N)}$ and
$B_3^{(N-1)}$ as a function of $B_2/B_3^{(N-1)}$. Curves labelled
as in fig.3~. } \label{fig4}
\end{figure}

In figure (\ref{fig2}) the results for the imaginary part of the
resonance complex energy are shown as a function of $B_2$ in units
of $\mu=1$. The solid, dashed and dotted lines are, respectively,
the correspondent imaginary part of $E_3^{(0)}$, $E_3^{(1)}$ and
$E_3^{(2)}$. The threshold values of $B_2$ for which the resonant
state becomes bound are clearly seen in the figure and the
resonance width, $\Gamma_3^{(N)}=2|Im[E_3^{(N)}]|$, increases with
the virtual dimer energy.
The results shown in figures (\ref{fig1}) and (\ref{fig2}) give
the complete trajectory of a three-body bound Efimov state when
the two-atom interaction with $a<0$ is changed, as can be done for trapped
atoms near a Feshbach resonance. These results extend the findings
of Ref.~\cite{yama02}. Now, we can give a complete picture of the
route of an Efimov state when $a$ is varied passing through a
Feshbach resonance. If one begins with positive $a$ and increases
it, a virtual trimer state becomes bound, then crossing a Feshbach
resonance and decreasing the magnitude of the large negative value
of $a$, the trimer bound state becomes unbound and turns into a
resonance.
The results for the trimer bound state or resonance energies can
be expressed in the form of a universal scaling function,
Eq.~(\ref{oe}), that depends only on the ratio of the two- and the
reference three-body energies. From the calculations given in
figures (\ref{fig1}) and (\ref{fig2}), we construct the scaling
plots of figures (\ref{fig3}) and (\ref{fig4}), respectively. For
this purpose, we use the first and second Efimov states (bound or
resonant) and we numerically obtain
\begin{equation}
E^{(N)}_{3} = B^{(N-1)}_{3}~{\cal
E}\left(-\sqrt{B_2/B^{(N-1)}_{3}}\right) \ , \label{oep}
\end{equation}
using $N$ equal to 1 and 2. (In Eq.~(\ref{oep}) the states are in
fact indexed, which was not explicitly shown in Eq.~(\ref{oe})).
We show the results for the real and imaginary part of the
resonance energy in a form of scaling plots in figures
(\ref{fig3}) and (\ref{fig4}), respectively. In these figures the
solid line corresponds to $N=1$ and the dotted line to $N=2$. In
figure $(\ref{fig3})$, the negative values of
$[Re(E_3)/B_3]^{1/2}$ give the trimer bound state results, while
the positive values come from the resonant state. It is clear that
a universal curve for the scaling function ${\cal
E}(-\sqrt{B_2/B_3})$ is approached as the results for $N=1$ and 2
practically overlap. The imaginary part of the resonance energy in
figure (\ref{fig4}) appears for the dimer virtual state energy
$B_2\geq 0.000882~B_3$, where the excited Efimov bound state turns
into the resonance. Our results extends the scaling function
${\cal E}(\sqrt{B_2/B_3})$ (bound dimer states) obtained
previously in \cite{am99} and \cite{yama02} to negative values of
the argument and for the trimer resonance region.

In conclusion, we have obtained the trajectory of an Efimov state
when the two-atom interaction crosses a Feshbach resonance, in a
form of a scaling function. The value of the resonance energy and
width depends only on the dimer bound or virtual energy and the
shallowest trimer binding energy. The route of an Efimov state
when $a$ is varied passing through a Feshbach resonance can be
summarized as follows: beginning from a large positive $a$ and
increasing it further, a virtual trimer state becomes bound and
then after crossing a Feshbach resonance and decreasing the
magnitude of the large negative value of $a$, the trimer bound
state becomes unbound and turns into a resonance, for
$a^{-1}<-0.0297\sqrt{m~B_3/\hbar^2}$, where $B_3$ is the binding
energy of the shallowest trimer state. In this respect, the study
of trapped Bose-Einstein condensates near a Feshbach resonance can
be fruitful not only to reveal the properties of an interacting
coherent quantum state under extreme conditions, but could also
give new insights in the few-body physics underlying the curious
phenomenon of Efimov resonant states.

We would like to thank Prof. Y. Koike from Hosei University for a
helpful discussion. We thank the Brazilian agencies FAPESP
(Funda\c c\~ao de Amparo a Pesquisa do Estado de S\~ao Paulo) and
CNPq (Conselho Nacional de Desenvolvimento Cient\'\i fico e
Tecnol\'ogico) for financial support.

\end{document}